\documentstyle[11pt,newpasp,twoside,epsf]{article}
\def\edcomment#1{\iffalse\marginpar{\raggedright\sl#1\/}\else\relax\fi}
\reversemarginpar

\begin{document}
\title{Star Formation from Turbulent Fragmentation }
 \author{Ralf S.\ Klessen}
\affil{UCO/Lick Observatory, University of California, 
 Santa Cruz, CA 95064, U.S.A.\\ Max-Planck-Institut f{\"u}r
 Astronomie, K{\"o}nigstuhl 17, 69117 Heidelberg, Germany}

\begin{abstract}
  Star formation is intimately linked to the dynamical evolution of
  molecular clouds. Turbulent fragmentation determines where and when
  protostellar cores form, and how they contract and grow in mass via
  accretion from the surrounding cloud material. Using numerical
  models of self-gravitating supersonic turbulence, efficiency,
  spatial distribution and timescale of star formation in turbulent
  interstellar clouds are estimated.  Turbulence that is not
  continuously replenished or that is driven on large scales leads to
  a rapid formation of stars in a clustered mode, whereas interstellar
  turbulence that carries most energy on small scales results in
  isolated star formation with low efficiency. The clump mass spectrum
  for models of pure hydrodynamic turbulence is steeper than the
  observed one, but gets close to it when gravity is included. The
  mass spectrum of dense cores is log-normal for decaying and
  large-wavelength turbulence, similar to the IMF, but is too flat in
  the case of small-scale turbulence. The three-dimensional models of
  molecular cloud fragmentation can be combined with dynamical
  pre-main sequence stellar evolution calculations to obtain a
  consistent description of all phases of the star formation
  process. First results are reported for a one solar mass protostar.
\end{abstract}

\section{Introduction}
\label{sec:intro}
Stars are born in interstellar clouds of molecular hydrogen, and
form by gravitational collapse of shock-compressed density
fluctuations generated from the supersonic turbulence ubiquitously
observed in molecular clouds (e.g.\ Elmegreen 1993, Padoan 1995,
Klessen, Heitsch, \& Mac~Low 2000).  Once a gas clump becomes
gravitationally unstable, collapse sets in and the central density
increases until a protostellar object forms which continues to grow in
mass via accretion from the infalling envelope.  As stars typically
form in small aggregates or larger clusters (Lada 1992, Adams \& Myers
2001) the interaction of protostellar cores and their competition for
mass growth from their surroundings are important processes shaping
the distribution of the final star properties.

Altogether star formation can be seen as a two-phase process: First,
{\em turbulent fragmentation} leads to transient clumpy molecular
cloud structure, with some of the density fluctuation exceeding the
critical mass and density for gravitational contraction. This sets the
stage for the second phase of star formation, the {\em collapse of
individual protostellar clumps} building up central protostars.  To
fully understand the star formation process and all its phases in
detail is one of the fundamental challenges in astronomy.  Describing
the first phase of star formation requires the three-dimensional
dynamical modeling of molecular cloud formation, evolution and
fragmentation. In this contribution, I report results from numerical
simulations using SPH, a particle-based approach to solve the
equations of hydrodynamics. The method and its application to (driven
or decaying) interstellar turbulence is introduced in \S{2} and \S{3}.
I demonstrate how supersonic turbulence may influence the properties
of forming (proto)stellar aggregates and clusters, and discuss the
spatial distribution and timescale of star formation in \S{4} and the
expected mass spectra for gas clumps and protostars in \S{5}. Finally,
in \S{6} a first attempt to combine {\em all} phases of star formation
into a consistent numerical scheme is presented.

\section{SPH --  A Particle-Based Flexible Method to Hydrodynamics}
\label{sec:SPH}
To adequately describe turbulent fragmentation and the formation of
protostellar cores, it is necessary to resolve the collapse of shock
compressed regions over several orders of magnitude in density. Due to
the stochastic nature of supersonic turbulence, it is not known in
advance where and when local collapse occurs. Hence, SPH ({\em
  smoothed particle hydrodynamics}) is used to solve the equations of
hydrodynamics. It is a Lagrangian method, where the fluid is
represented by an ensemble of particles and thermodynamical
observables are obtained by averaging over an appropriate subset of
the SPH particles (Benz 1990). High density contrasts are resolved by
simply increasing the particle concentration where needed.  SPH can
also be combined with the special-purpose hardware device GRAPE
(Sugimoto et al.\ 1990, Ebisuzaki et al.\ 1993, Steinmetz 1996)
allowing for calculations at supercomputer level on a normal
workstation. The simulations presented here concentrate on subregions
within a much larger cloud, therefore periodic boundary conditions are
adopted (Klessen 1997). The high-density cores of collapsing gas
clumps are  substituted by  `sink' particles
(Bate, Bonnell \& Price 1995) while keeping track of mass and
 momentum. This allows for following the dynamical evolution of the
system over many free-fall timescales.
 
\section{Models of Driven Supersonic Turbulence}
\label{sec:turb}
 The large observed linewidths in molecular clouds imply the presence
 of supersonic velocity fields that carry enough energy to
 counterbalance gravity on global scales (Williams, Blitz, \& McKee
 2000).  However, it is known that turbulent energy dissipates
 rapidly, roughly on the free-fall timescale (Mac Low et al.\ 1998,
 Stone, Ostriker, \& Gammie 1998, Padoan \& Nordlund 1999). Unlike
 previously thought, this is independent of the presence of magnetic
 fields. The fields are therefore taken as being dynamically
 unimportant in the current models, and are not included. To prevent
 or considerably postpone global collapse, turbulence must be
 continuously replenished. This is achieved by applying a Gaussian
 driving scheme, that inserts kinetic energy in a specified range of
 wavenumbers $k$. I select $1 \le k \le 2$, $3 \le k \le 4$, and $8 \le k
 \le 8$, corresponding to sources that act on large, intermediate, and
 small scales, respectively.  The energy input at each timestep is
 adjusted to reach a constant level of the total kinetic energy
 sufficient to stabilize the cloud as a whole (Mac~Low 1999).  Furthermore one
 molecular cloud model is considered where turbulence is assumed to
 have decayed, leaving behind Gaussian density fluctuations which
 begin to contract on all scales. For details see Klessen (2001).

The models presented here are computed in normalized units. If scaled
to mean densities of $n({\rm H}_2) = 10^4\,$cm$^{-3}$, a value typical
for star-forming molecular cloud regions (e.g.\ in $\rho$-Ophiuchus,
see Motte, Andr{\'e}, \& Neri 1998) and a temperature of 11.4$\,$K
(i.e.\ a sound speed $c_{\rm s} = 0.2\,$km/s), then the total mass
contained in the computed volume is 413$\,$M$_{\odot}$ and the size of
the cube is $0.89\,$pc. It contains 64 thermal Jeans masses.

\section{Spatial Distribution and Timescale of Star Formation}
\label{sec:location-time}
Stars form from turbulent fragmentation of molecular cloud material.
Supersonic turbulence that is strong enough to counterbalance
gravity on global scales will usually provoke {\em local} collapse.
Turbulence establishes a complex network of interacting shocks, where
converging shockfronts  generate clumps of high-density. This density
enhancement can be large enough for the fluctuations to become
gravitationally unstable and collapse, i.e.\ when the local Jeans
length becomes smaller than the size of the fluctuation.  However, the
fluctuations in turbulent velocity fields are highly transient.  The
random flow that creates local density enhancements can disperse them
again.  For local collapse to actually result in the formation of
stars, locally Jeans unstable, shock-generated, density fluctuations
must collapse to sufficiently high densities on time scales shorter
than the typical time interval between two successive shock passages.
Only then are they able to `decouple' from the ambient flow pattern
and survive subsequent shock interactions.  The shorter the time
between shock passages, the less likely these fluctuations are to
survive. Hence, the efficiency of protostellar core formation and the
rate of continuing accretion onto collapsed cores depend strongly on
the wavelength and strength of the driving source
(see Klessen et al.\ 2000, Heitsch, Mac~Low, \& Klessen 2001).

The velocity field of long-wavelength turbulence is dominated by
large-scale shocks which are very efficient in sweeping up molecular
cloud material, thus creating massive coherent structures. When a
coherent region reaches the critical density for gravitational collapse
its mass typically exceeds the local Jeans limit by far.  Inside the
shock compressed region, the velocity dispersion is much smaller than
in the ambient turbulent flow and the situation is similar to
localized tur\-bulent decay. Quickly a cluster of protostellar cores
builds up. Both, decaying and large-scale turbulence lead to a {\em
  clustered} mode of star formation. The efficiency of turbulent
fragmentation is reduced if the driving wavelength decreases. When
energy is carried mainly on small spatial scales, the network of
interacting shocks is very tightly knit, and protostellar cores form
independently of each other at random locations throughout the cloud
and at random times.  Individual shock generated clumps have lower
mass and the time interval between two shock passages through the same
point in space is small.  Hence, collapsing cores are easily destroyed
again and star formation is inefficient. This scenario corresponds to
the {\em isolated} mode of star formation. It needs to be pointed out
that there is no fundamental dichotomy between the two modes of star
formation, they rather define the extreme ends in the continuous
spectrum of the properties of turbulent molecular cloud fragmentation.

\begin{figure*}[t]
\unitlength1cm
\begin{picture}(15.0, 7.0)
\put(-0.3, 0.0) {\epsfxsize=7.0cm \epsfbox{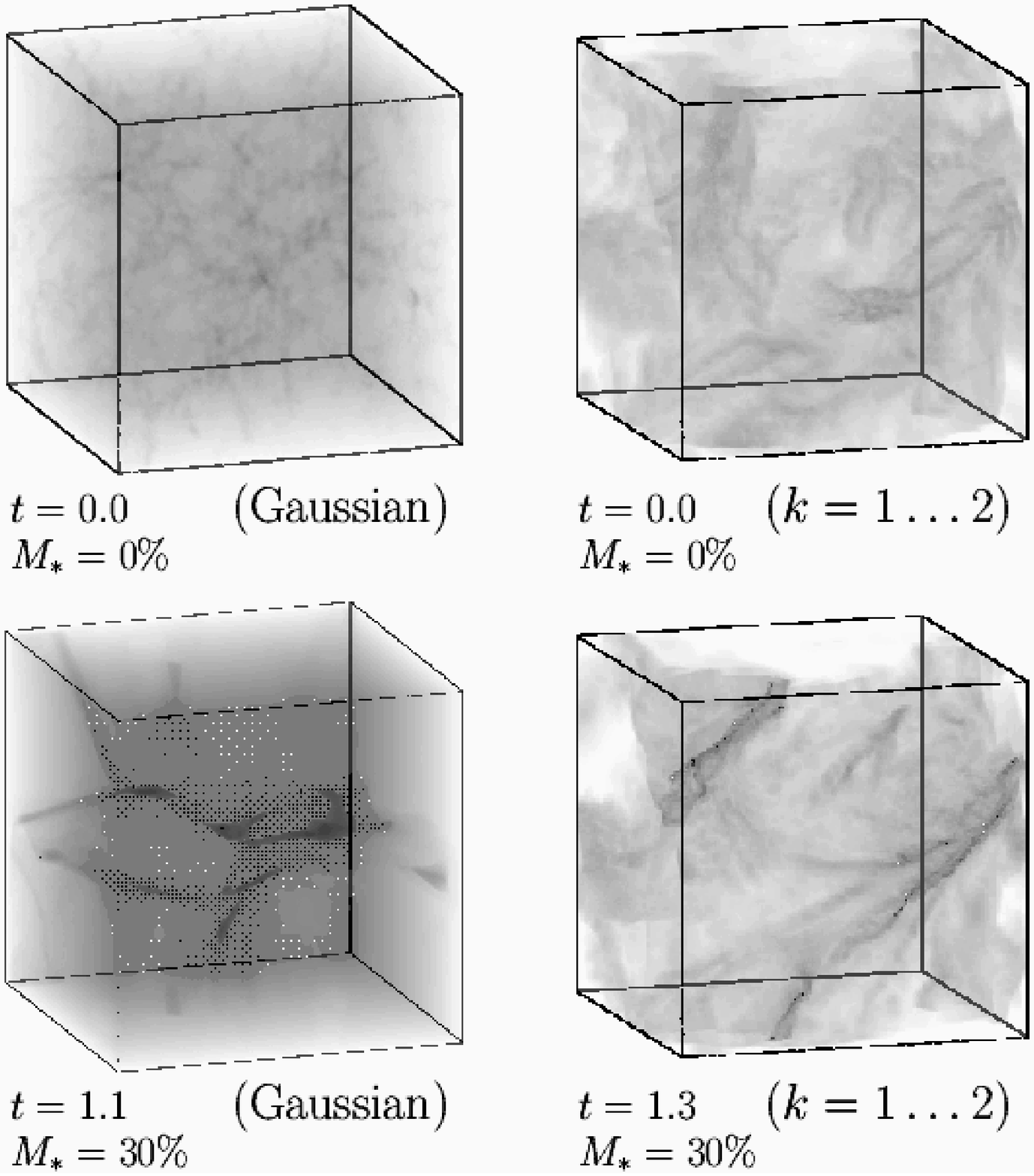}}
\put( 6.4, 0.0) {\epsfxsize=7.0cm \epsfbox{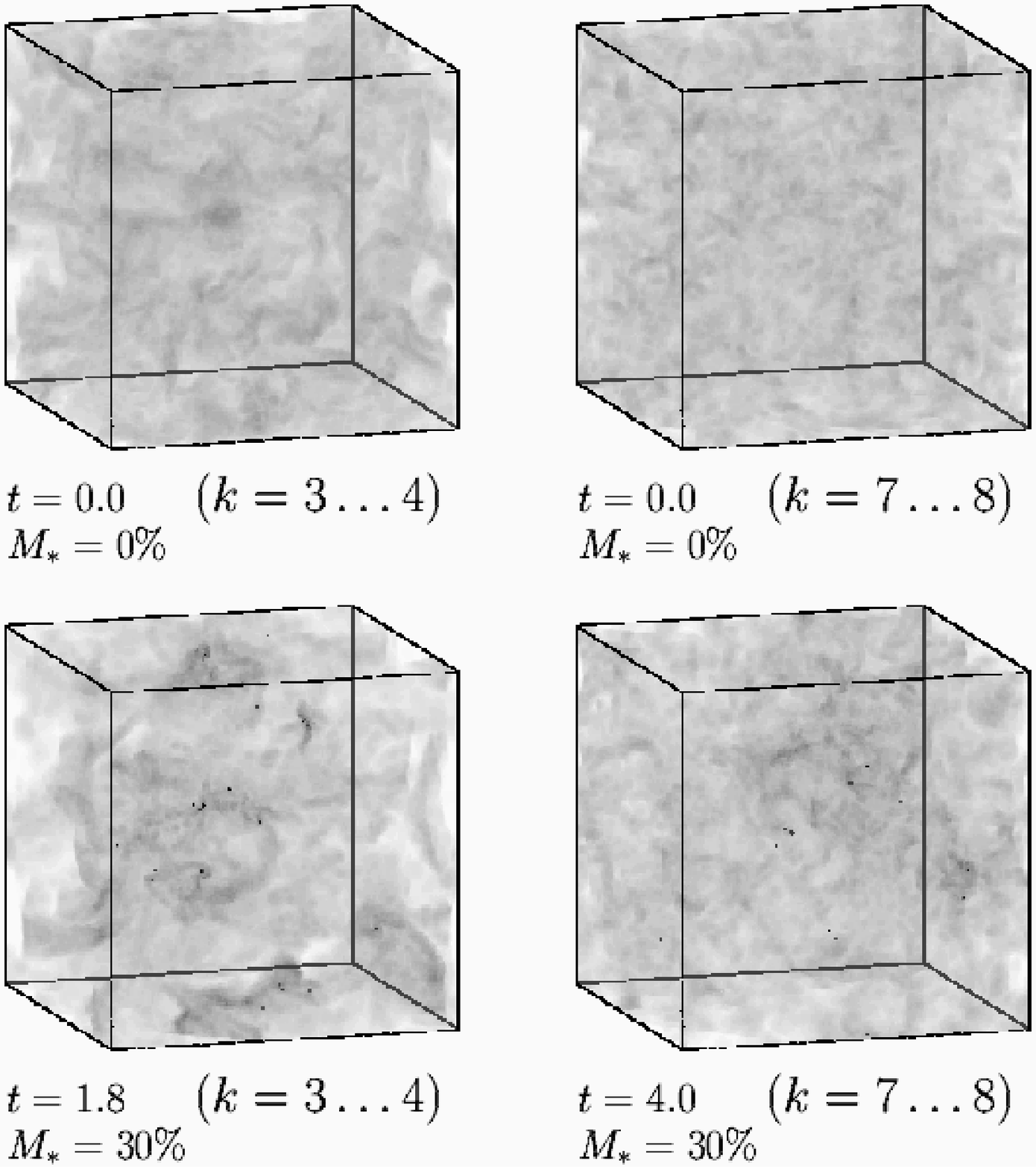}}
\end{picture}
\caption{Comparison of the gas distribution in the four models:
decayed turbulence leading to a Gaussian field (left column),
large-scale turbulence (second column), intermediate-wavelength
turbulence (third column), and small-wavelength turbulence (right
column).  The upper panel depicts the initial stage when gravity is
`turned on', the lower panel shows system after the first cores have
formed and accumulated roughly 30\% of the total mass.
%
%
  }
\end{figure*}
%

\begin{figure*}[t]
\unitlength1cm
\begin{picture}(15.0, 8.5)
\put( 1.4, 0.0) {\epsfxsize=11.0cm \epsfbox{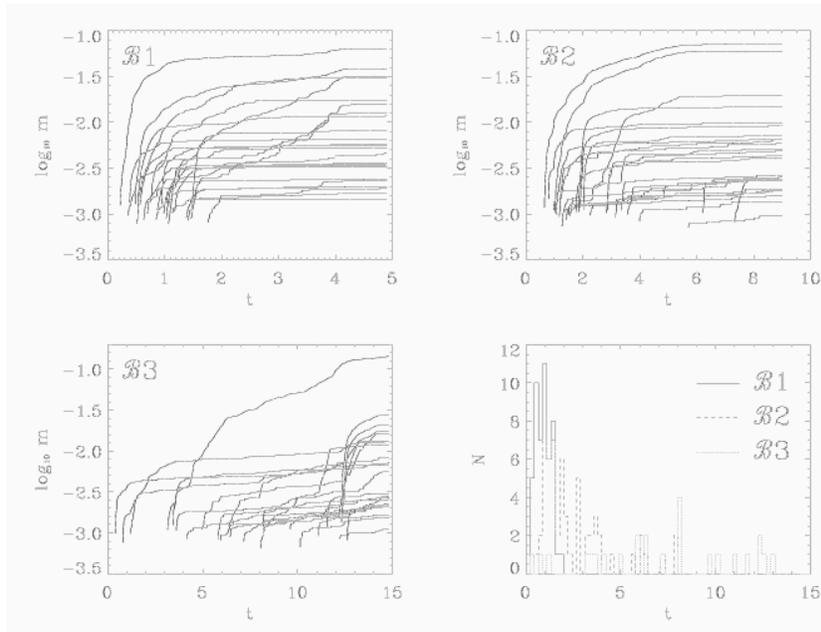}}
\end{picture}
\caption{Mass growth history of protostellar cores (only every
  second core is depicted): long-wavelength (${\cal B}1h$),
  intermediate- (${\cal B}2h$), and small-scale turbulence (${\cal
    B}3h$). The lower left plot shows the distribution of the
  formation times of the cores. Time is  given in units of the
free-fall time 
  and masses are scaled to the total mass in the system.  }
\end{figure*}
This is visualized in Fig.~1. It compares the models of decayed,
large-wavelength ($k=1\dots2$), intermediate ($k=3\dots4$), and
small-scale turbulence ($k=7\dots8$). The density structure of the
systems is depicted at $t=0$, which means for the model of decayed
turbulence that the initial Gaussian density field is visible. For the
driven models, $t=0$ corresponds to the phase of fully developed
turbulence just before gravity is `switched on'. The lower panel
describes the four models after the first protostellar cores have
formed via turbulent fragmentation and have accreted 30\% of the total
mass.  Time is measured in units of the global  free-fall
timescale. Dark dots indicate the
location of dense collapsed core. As for decayed turbulence all
spatial modes are unstable, the system quickly evolves into a network
of intersecting filaments, where protostellar cores predominantly form.
Similarly, also large-scale turbulence builds up a network of
filaments, however, this time the large coherent structures are not
caused by gravity, but instead are due to shock compression.  Once
gravity is included, it quickly dominates the evolution inside the
dense regions and again a cluster of protostellar cores builds
up. In the case of intermediate-wavelength turbulence, cores form in
small aggregates, whereas small-scale turbulence leads to local
collapse of individual objects randomly dispersed throughout the
volume. Note the different times needed for 30\% of the mass to be
accumulated in dense cores.  For small-scale turbulence star formation
needs longest and speeds up with increasing wavelength, and as
expected the rates for large-scale turbulence and locally decayed
turbulence are comparable.  This is also seen in Fig.~2, which shows
the mass accretion history of individual protostellar cores for the
three turbulent models, together with the distribution of core
formation times. Star formation efficiency is high in the
long-wavelength model, all cores form within two free-fall times,
whereas in the short-wavelength model the efficiency is low and core
formation continues for over 15 free-fall timescales (when the simulation
was stopped).

\section{Mass Spectra of Clumps and Protostellar Cores}
\label{sec:mass-spectra}
The dominant parameter determining stellar evolution is the mass. It
is therefore important to investigate the relation between the masses
of molecular clumps, protostellar cores and the resulting stars.
Figure 3 plots for the four models the mass distribution of all gas
clumps, of the subset of Jeans critical clumps, and of collapsed
cores. Four different evolutionary phases are shown, initially just
when gravity is `switched on', and after turbulent fragmentation has
lead to the accumulation of $M_{\rm \large *}\approx 5$\%, $M_{\rm
\large *}\approx 30$\% and $M_{\rm \large *}\approx 60$\% of the total
mass in dense cores.

In the initial, completely pre-stellar phase the clump mass spectrum
is  very steep (about Salpeter slope or less) at the high-mass end and
gets shallower below $M \approx 0.4 \,\langle M_{\rm J} \rangle$ with
slope $-1.5$, when using a power-law fit. The spectrum strongly declines
beyond the SPH resolution limit. Altogether, individual clumps are
hardly more massive than a few $\langle M_{\rm J}\rangle$.

\begin{figure*}[th]
\unitlength1cm
\begin{picture}(15.0,10.4)
\put( 0.5, -0.5) {\epsfxsize=12.0cm \epsfbox{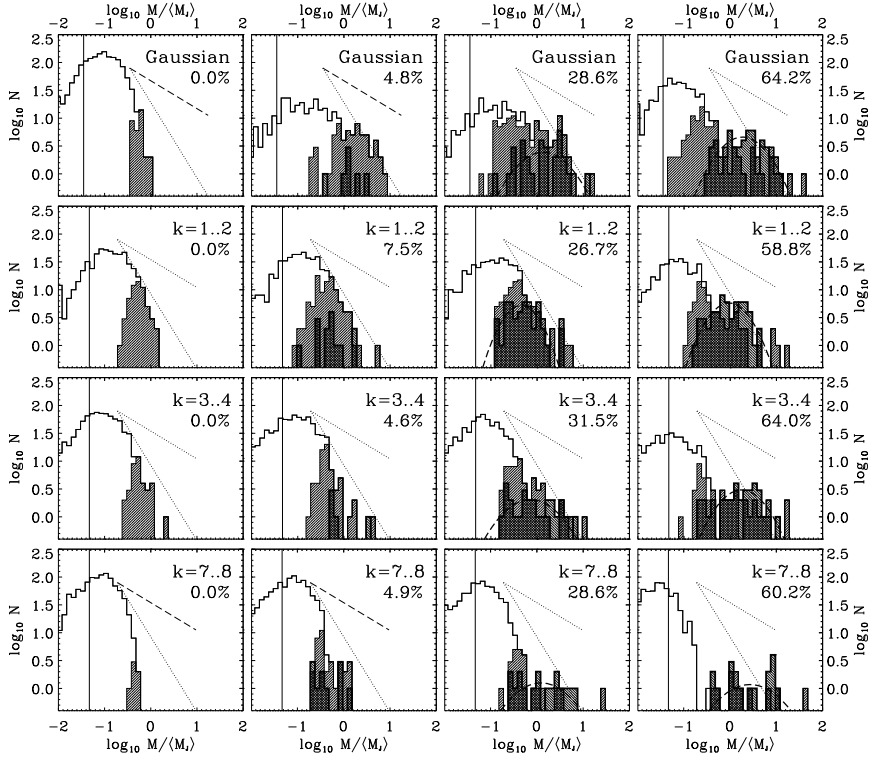}}
\end{picture}
\label{fig:mass-spectrum}
\caption{Mass spectra of dense collapsed cores (hatched thick-lined
histograms), of gas clumps (thin lines), and of the subset of Jeans
unstable clumps (thin lines, hatched distribution).  Masses are binned
logarithmically and normalized to the average Jeans mass $\langle
M_{\rm J}\rangle$. The left column gives the initial state of the
system, just when gravity is `switched on', the second column shows
the mass spectra when roughly $M_{\rm \large *}\approx 5$\% of the
mass is in dense cores, the third column when this fraction is
about 30\%, and the right column when $M_{\rm \large *}\approx
60$\%. For comparison with power-law spectra ($dN/dM \propto
M^{\nu}$), the typical slope $\nu = -1.5$ of the observed clump mass
distribution, and the Salpeter slope $\nu=-2.33$ for the IMF, are
indicated by the long dashed and by the dotted lines in each plot.
The vertical line shows the SPH resolution limit.}
\end{figure*}
Gravity strongly modifies the distribution of clump masses during the
later evolution. As clumps merge and grow bigger, the mass spectrum
becomes flatter and extends towards larger masses. Consequently the
number of cores that exceed the Jeans limit grows, and local collapse
sets in leading to the formation of dense cores. This is most evident
in the Gaussian model of decayed turbulence, where the velocity field is
entirely determined by gravitational contraction on all scales. The
clump mass spectrum in intermediate phases of the evolution (i.e.\ 
when protostellar cores are forming but the overall gravitational
potential is still dominated by non-accreted gas) exhibits a slope
$-1.5$ similar to the observed one. When the velocity field is
dominated by strong (driven) turbulence, the effect of gravity on the
clump mass spectrum is much weaker. It remains steep, close to or even
below the Salpeter value. This is seen for
small-wavelength turbulence. Here, the short interval between shock
passages prohibits efficient merging and the build up of a large
number of massive clumps. Only few clumps become Jeans unstable and
collapse to form cores. These form independent of each other at random
locations and times and typically do not interact.  Increasing the
driving wavelength leads to more coherent and rapid core formation,
which also results in a larger number of cores. It is
maximum in the case of pure gravitational contraction (decayed
turbulence).

Long-wavelength turbulence or turbulent decay leads to a core mass
spectrum that is well approximated by a {\em log-normal}. It roughly
peaks at the {\em average thermal Jeans mass} $\langle M_{\rm
  J}\rangle$ of the system and is comparable in width with the
observed IMF (see Klessen \& Burkert 2000, 2001). The log-normal shape of
the mass distribution may be explained by invoking the central limit
theorem (e.g.\ Zinnecker 1984), as protostellar cores form and evolve
through a sequence of highly stochastic events (resulting from
supersonic turbulence and/or competitive accretion). To find the mass
peak at $\langle M_{\rm J}\rangle$ may be somewhat surprising given
the fact that the local Jeans mass strongly varies between different
clumps. In a statistical sense the system retains knowledge of its
mean properties.  The total width of the core distribution is about
two orders of magnitude in mass and is approximately the same for all
four models.  However, the spectrum for intermediate and
short-wavelength turbulence, i.e.\ for isolated core formation, is too
flat (or equivalently too wide) to be comparable to the observed IMF.
This is in agreement with the hypothesis that most stars form in
aggregates or clusters.

\begin{figure*}[tp]
\unitlength1cm
\begin{picture}(16.0,8.8)
\put( -0.2, 0.0){\epsfxsize=12.8cm \epsfbox{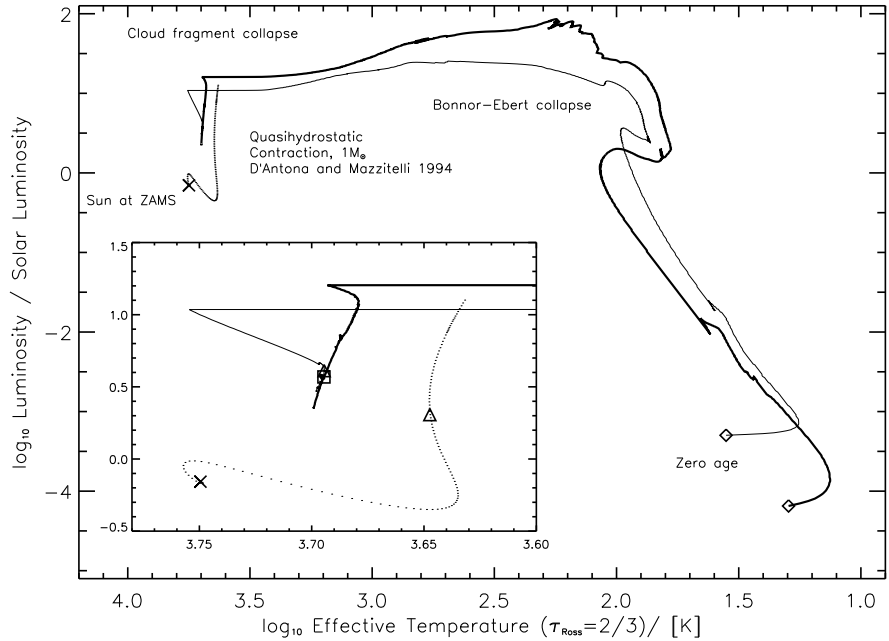}}
\end{picture}
\caption{ Early stellar evolution in the Hertzsprung-Russell diagram.
Three evolutionary effective-temperature-luminosity relations (tracks)
relevant to the young Sun are compared. The dotted line is a classical
stellar structure, hydrostatic-equilibrium PMS-track for $1\,{\rm
M_\odot} $, for an initially fully convective gas sphere (`MLT
Alexander' model of D'Antona \& Mazzitelli 1984). The two other lines
are obtained by describing the formation of the star as a result of
the collapse of an interstellar cloud.  The {\em thin line} is for a
cloud fragment in initial equilibrium (a so called `Bonnor-Ebert'
sphere of a solar mass, see Wuchterl \& Tscharnuter 2001 for
details). The (thick line) is for a cloud fragment that results from
the dynamical fragmentation of a molecular cloud (model $\cal I$ of
Klessen \& Burkert 2000, see Wuchterl \& Klessen 2001 for details).
The two diamonds, in the lower right, indicate zero age for the two
collapse-results, when the protostellar fragments for the first time
become optically thick and depart from isothermality.  Triangles and
squares mark the point along the respective evolutionary tracks where
an age of 1 million years is reached.  The cross on the hydrostatic
track denotes the moment when energy generation by nuclear reactions
in the stellar interior, for the first time in stellar live {\em
fully} compensates the energy losses due to radiation from the stellar
photosphere, i.e. the zero age main sequence (ZAMS). Note, that once
accretion fades away the dynamical PMS calculacions converge and
predict that the proto-Sun was $500\,$K hotter and twice as luminous
as in the classical hydrostatic model. The figure is a courtesy of G.\
Wuchterl.  }
\end{figure*}
\section{Towards an Integral Description of Star Formation}
\label{sec:SF}
Three-dimensional calculations of molecular cloud dynamics can only
describe the first phases of the star formation process: turbulent
molecular cloud fragmentation and formation of collapsing protostellar
cores. These calculation fail to resolve the build-up of the protostar
itself. Understanding the temporal evolution of its structural
parameters and the determination of observables like luminosity and
effective temperature as the star approaches the main sequence
requires detailed protostellar evolutionary calculations. These
calculations are important, because they give ages, masses and radii
of young stars when brightness, distance and effective temperature are
known.  The so determined ages constitute the only practical `clock'
for tracing the history of star-formation regions and for studying the
evolution of circumstellar disks and planet formation.

However, almost all classical PMS calculations start {\em after} the main
protostellar accretion phase ended. Formation and mass growth of the
protostar are usually neglected, and its internal thermal structure is
taken to be fully convective (e.g.\ D'Antona \& Mazzitelli 1994).
However, it has been shown recently that protostellar collapse does
{\em not} lead to a fully convective stellar structure when calculated
consistently, taking into account radiation-hydrodynamics,
time-dependent convection and nuclear burning processes (Wuchterl \&
Tscharnuter 2001). 

A fully {\em consistent} treatment of formation and
evolution of protostars requires  coupling the
 3D cloud fragmentation models with
the  1D dynamical description of protostellar
collapse.  This can be achieved by taking the  physical
parameters and the time dependent mass accretion rates onto individual
protostellar cores from the 3D model  as boundary values for
the detailed 1D calculation (see Wuchterl \& Klessen 2001). 
For the proto-Sun, i.e.\ a cloud fragment
with $1\,$M$_{\odot}$, the evolutionary path in the
Hertzsprung-Russell (HR) diagram is illustrated in Fig.~4. 

These calculations demonstrate that luminosity and temperature of
protostars not only depend on mass and age, but also on the accretion
rate in the main accretion phase. Therefore, the position of a
protostar on the HR diagram is sensitive to the cluster environment,
because stochastic core interaction and competition strongly influence
the mass accretion rate: {\em PMS tracks are not unique, and mass (or
similarly age) of individual protostars from using the HR diagram can
only be determined in a statistical sense.}

Furthermore, the new dynamical models predict considerable differences
to the currently established classical PMS tracks: Instead of being
fully convective, the structure of a $1\,$M$_{\odot}$ protostar is
homologous to the present day Sun.  Already an age of 1 million years,
it has a radiative core and a convective envelope. For this reason,
the dynamical PMS calculacions predict that the proto-Sun was $500\,$K
hotter and twice as luminous as in the classical hydrostatic model.
Altogether, conclusive observational tests are necessary to determine
the validity of the numerical models. It is essential in this context
to obtain independent mass and/or age determinations, e.g.\ from PMS
binary stars or detailed modeling of protostellar disks (see Simon,
Dutrey, \& Guilloteau 2000, or Woitas, K{\"o}hler, \& Leinert 2001 for
first estimates).

\acknowledgements I thank G{\"u}nther Wuchterl, Mordecai-Mark Mac~Low,
 Fabian Heitsch, and Peter Bodenheimer for many stimulating
 discussions and fruitful collaboration.

\end{document}